\documentclass[cits]{PoS}
\let\ifpdf\relax
\usepackage[authoryear,square]{natbib}
\bibpunct{(}{)}{;}{a}{}{,}
\usepackage{graphicx} 
\usepackage{amssymb}
\usepackage{fontenc}
\usepackage{times} 
\usepackage{mathptmx}
\usepackage{sidecap}

\title{Non-thermal emission from galaxy clusters: feasibility study with SKA1}

\ShortTitle{SKA1: feasibility study for cluster non-thermal emission}

\author{
\speaker{Chiara Ferrari}$^1$, 
Arwa Dabbech$^1$, 
Oleg Smirnov$^{2, 3}$, 
Sphesihle Makhathini$^2$, 
Jonathan S. Kenyon$^2$, 
Matteo Murgia$^4$, 
Federica Govoni$^4$, 
David Mary$^1$, 
Eric Slezak$^1$, 
Franco Vazza$^5$, 
Annalisa Bonafede$^5$, 
Marcus Br\"uggen$^5$, 
Melanie Johnston-Hollitt$^6$, 
Siamak Dehghan$^6$, 
Luigina Feretti$^7$, 
Gabriele Giovannini$^{7, 8}$, 
Valentina Vacca$^9$, 
Michael Wise$^{10}$,
Myriam Gitti$^{7, 8}$, 
Monique Arnaud$^{11}$, 
Gabriel W. Pratt$^{11}$, 
Kristian Zarb Adami$^{12, 13}$,
Sergio Colafrancesco$^{14}$
\\
$^1$ Laboratoire Lagrange, UMR 7293, UNS, CNRS, OCA, 06300 Nice (FR) \\
$^2$ Department of Physics and Electronics, Rhodes University, PO Box 94, Grahamstown, 6140 \\
$^3$ SKA South Africa, 3rd Floor, The Park, Park Road, Pinelands, 7405 (ZA) \\
$^4$ INAF - Osservatorio Astronomico di Cagliari, Via della Scienza 5, 09047 Selargius (IT) \\
$^5$ Universit\"at Hamburg, Hamburger Sternwarte, Gojenbergsweg 112, D-21029, Hamburg (DE) \\
$^6$ School of Chemical \& Physical Sciences, Victoria University of Wellington, PO Box 600, Wellington, 6140 (NZ)\\
$^7$ INAF -- Istituto di Radioastronomia, Via Gobetti 101, I--40129 Bologna (IT)\\
$^8$ Dipartimento di Fisica e Astronomia, Universit\`a di Bologna, Via Ranzani 1, I--40127 Bologna (IT) \\
$^9$ Max Planck Institute for Astrophysics, Karl-Schwarzschild-Str. 1, 85748 Garching (DE)\\
$^{10}$ Netherlands Institute for Radio Astronomy (ASTRON),
Postbus 2, 7990 AA Dwingeloo (NL)\\
$^{11}$ Laboratoire AIM, IRFU/Service d'Astrophysique -- CEA F--91191 Gif-sur-Yvette Cedex (FR)\\
$^{12}$ Physics Department, University of Malta, Msida, MSD 2080 (MT) \\
$^{13}$ Physics Department, University of Oxford, Oxford, OX1 3RH (UK)\\
$^{14}$ School of Physics, University of the Witwatersrand, Private Bag 3, 2050-Johannesburg (ZA)\\
E-mail: \email{chiara.ferrari at oca.eu}
}

\abstract{Galaxy clusters are known to host a variety of extended radio sources: tailed radio galaxies whose shape is modelled by the interaction with the intra-cluster medium (ICM); radio bubbles filling cavities in the ICM distribution and rising buoyantly through the thermal gas; diffuse giant radio sources (``halos'' and ``relics'') revealing the presence of relativistic electrons and magnetic fields in the intra-cluster volume. It is currently the subject of an active debate how the non-thermal components that we observe at radio wavelengths affect the physical properties of the ICM and depend on the dynamical state of galaxy clusters.

In this work we start our SKA1 feasibility study of the ``radio cluster zoo'' through simulations of a typical radio-loud cluster, hosting several bright tailed radio galaxies and a diffuse radio halo. Realistic simulations of SKA1 observations are obtained through the {\tt MeqTrees} software. A new deconvolution algorithm, based on sparse representations and optimised for the detection of faint diffuse astronomical sources, is tested and compared to the classical {\tt CLEAN} method.  }

\FullConference{
Advancing Astrophysics with the Square Kilometre Array\\
June 8-13, 2014\\
Giardini Naxos, Sicily, Italy}

\newcommand{\skipthis}[1]{}

\begin{document}

\section{Science case}

The discovery of diffuse radio sources up to Mpc scales (called ``halos'', ``mini-halos'' or ``relics'', depending on their position in the cluster, size, morphology and polarization properties) in more than 70 galaxy clusters has pointed out the existence of a non-thermal (NT) component (relativistic electrons with Lorentz factor $>>$1000 and magnetic fields of the order of $\mu$G) in the intracluster volume (e.g. \cite{ferrari08}; see also Chapters by \citeauthor{cassano15,govoni15,gitti15}, this Volume). Through NT studies of galaxy clusters we can estimate the cosmic-ray and magnetic field energy budget and pressure contribution to the intracluster medium (ICM), as well as get clues about energy redistribution during cluster mergers. NT analyses can elucidate non-equilibrium physical processes whose deep understanding is essential to do high-precision cosmology using galaxy clusters \citep{vazza12a}.

A detailed understanding of the origin of the intracluster NT component is still missing. A current status of this research area is summarised here and is presented in more detail in the Chapter by R. Cassano et al. While magnetic fields have been proven to be ubiquitous in the intracluster volume \citep{bonafede11}, it is still debated how the thermal electrons of the ICM can be accelerated to relativistic energies. Since the radiative lifetime of electrons is much shorter than their crossing time over Mpc scales, cosmic ray acceleration has to be related to ``in situ'' physical processes. Diffuse radio emission has generally been detected in massive merging clusters. The most widely accepted acceleration models are thus those that predict electron acceleration by shocks (in the case of relics) and turbulence (in the case of halos) that develop within the ICM during cluster interactions \citep[e.g.][]{brunetti14}. Note, however, that recently \citet{bonafede14a} have pointed out the existence of a giant radio halo in a cluster characterised by a cool-core, i.e. either a nearly relaxed or a minor merger system. Relativistic electrons are also expected to be produced in clusters as a secondary product of hadronic collisions between the ions of the ICM and relativistic protons, characterised by significantly longer lifetime compared to relativistic electrons \citep[][and refs. therein]{ensslin11}. Even if most evidence indicates that secondary electrons are not expected to give rise to diffuse radio emission at levels detectable by current instruments \citep[but see][]{ensslin11}, they could  provide the seeds for further re-acceleration by merger induced turbulence and shocks. There is therefore still the need to disentangle their possible contribution to the total cluster radio emission through the next generation of radio telescopes \citep{brunetti11}. Theoretical models of electron acceleration need to be compared to statistical samples of clusters emitting at radio wavelengths, while only a few tens of radio relics and halos are known up to now and mostly at low/moderate redshift \citep[$z \lesssim 0.4$,][]{feretti12}. Of great importance for characterising the origin of intracluster cosmic rays is the possibility to perform spectral analyses of diffuse radio sources \citep[e.g.][]{orru07, stroe13}. While currently deep pointed radio observations have allowed to detect radio emission from a few cases of high-z clusters \citep[$z > 0.5$, see][and references therein]{vanweeren14}, it is crucial to perform radio studies of statistical cluster samples up to z$\sim$1 (to follow the assembly process from the epoch of massive cluster formation) and get detailed information about the mass and dynamical state of ``radio loud'' vs. ``radio quiet'' clusters \citep[e.g.][]{govoni04, cassano13}.

Apart from radio halos and relics, galaxy clusters host a wider variety of extended radio sources, such as tailed radio galaxies whose shape is modelled by the interaction with the ICM \citep[e.g.][]{dehghan11,pfrommer11,pratley13} and radio bubbles filling holes in the ICM distribution and rising buoyantly through the thermal gas \citep[e.g.][]{degasperin12, gitti12}. Joint studies of all types of extended radio sources in clusters allow us to address the complex physical processes regulating the interaction between the different components of galaxy clusters \citep[e.g.][]{bonafede14b}. For this, it is of course crucial to be able to identify separately the different kinds of radio sources in galaxy clusters (i.e. to discriminate between the radio emission related to active galaxies or to the NT ICM). In this paper we start our SKA1 feasibility study of the ``radio cluster zoo'' through a model radio-loud cluster presented in Sect.\,\ref{model}. Realistic simulations of SKA1 observations of this model cluster are described in Sect.\,\ref{simulations}. Sect.\,\ref{moresane} focuses on the results of a new deconvolution algorithm \citep[{\tt MORESANE},][]{dabbech12, dabbech14} optimised for the detection of faint diffuse astronomical sources. We conclude with some remarks about the feasibility of cluster studies with SKA1 and with future plans in Sect.\,\ref{conclusion}.

\section{Simulations of a radio-loud galaxy cluster} 
\subsection{The model cluster} \label{model}

\begin{SCfigure} 
\centering
\caption{Simulated radio emission at 1.4 GHz from a galaxy cluster \citep[{\tt FARADAY} tool,][]{murgia04} at z=0.5. The radio galaxy population is extracted from the cluster A\,2255 \citep{govoni06}.}
\includegraphics[width=0.5\columnwidth]{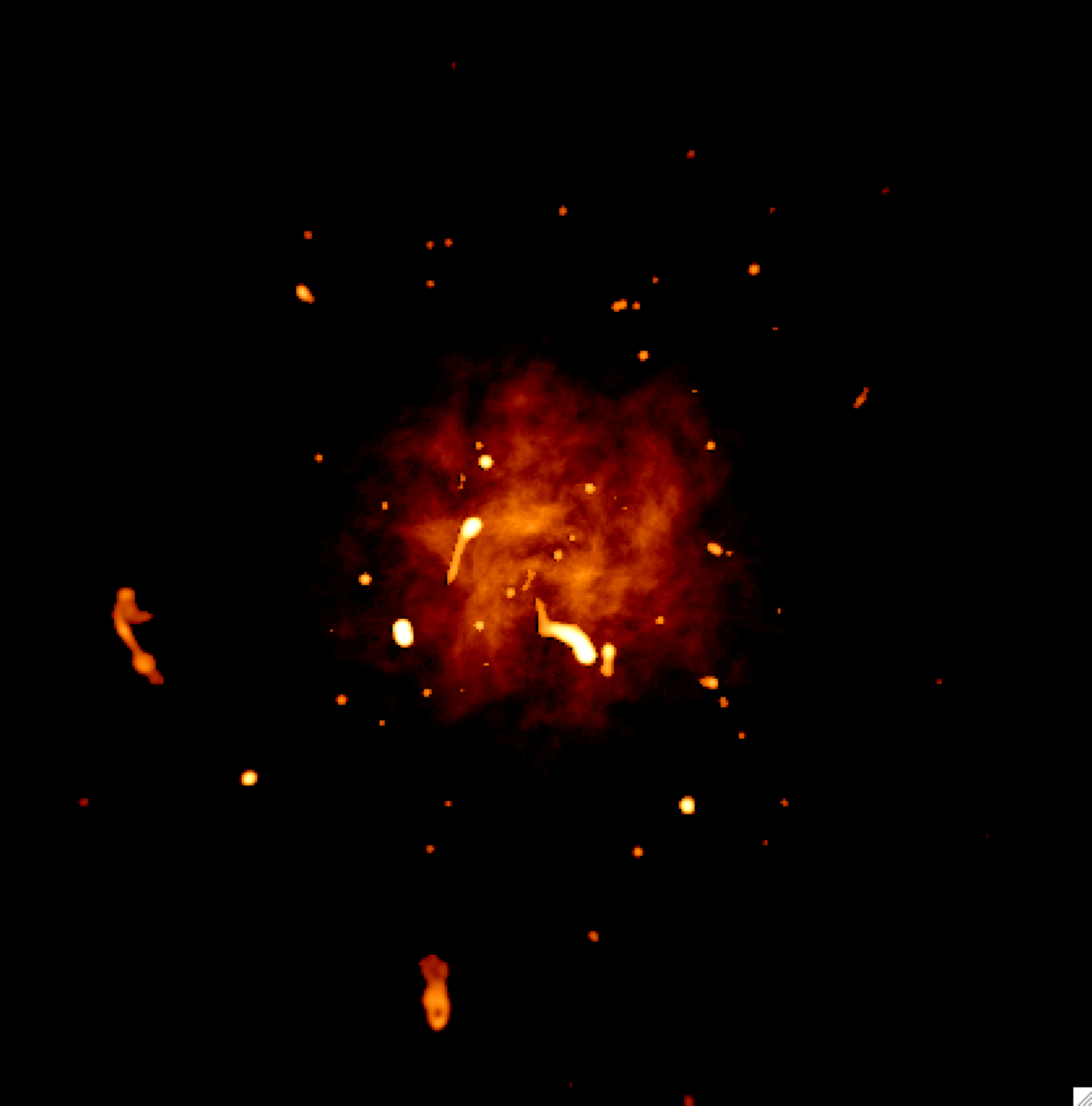}
\label{fig:model}
\end{SCfigure}

While with current facilities radio halos have been mostly discovered in low-redshift clusters \citep[$<{\rm z}> \sim 0.2$,][]{feretti12}, with this study we aim at analysing up to which redshift we can detect diffuse cluster radio emission with SKA1. We are particularly interested to test if we can reach the epoch in which massive clusters that we observe today are forming, i.e. z $\approx$ 1. By modelling both the gas and energy density distributions of the thermal and relativistic electron populations, and the characterisation of the magnetic field fluctuation and radial scaling similarly to \citet{govoni06}, we perform simulations of a galaxy cluster at z=0.5 using the {\tt FARADAY} tool \citep{murgia04}. The resulting model cluster hosts a diffuse radio halo, several tailed radio galaxies and point sources (see Fig.\,\ref{fig:model}). The total power of the simulated radio halo is $P_{\rm 1.4~GHz} \sim$ $1.2 \times 10^{24}$ W/Hz, roughly corresponding to the luminosity limit of currently detected radio halos (left panel of Fig.\,1 in \citeauthor{cassano15}, this Volume). The overlaid radio galaxy population is extracted from the galaxy cluster A\,2255 \citep{govoni06}. The simulated model cluster is then redshifted up to z=1.0 by taking into account the scaling of size, surface brightness and radio luminosity with redshift \citep[e.g.][]{ensslin02}.

\subsection{Simulations of SKA1-MID and SKA1-SUR observations} \label{simulations}

In this work, we start from our model cluster described in Sect.\,\ref{model} and we perform realistic simulations of SKA1-MID and SKA1-SUR observations using the {\tt MeqTrees} software \citep{noordam10}. We provide as input currently available antenna configurations and, similarly to the ``SKA1 imaging science performance'' document by R. Braun, a total observing time of 8h. Note that, based on the bigger field-of-view (FoV) of SKA-SUR with respect to SKA1-MID ($\sim$ 18 deg$^2$ {\it vs} 0.38 deg$^2$ at 1.4 GHz), the first instrument would approximately allow an all-sky survey within 2 years with the adopted observation time per field. Conversely, the higher sensitivity of SKA1-MID provides a significantly better detection of cluster radio emission, as discussed in the following.
In order to limit the simulated data volume, a 60s integration time is assumed and no time averaging is applied. A 50 MHz bandwidth of a single channel and starting at 1415 MHz is considered. The simulated observations are treated as essentially monochromatic. No primary beam corrections are applied: the size of the input model map is selected to be 2048 $\times$ 2048 pixels$^2$, with the 512 $\times$ 512 pixels$^2$ sky image shown in Fig.\,\ref{fig:model} padded with zeros in its external regions. We use SEFDs\footnote{System Equivalent Flux Densities} as set out in the baseline design \citep{dewdney13} and rescale the noise to simulate the 8 hours of synthesis.

\begin{figure*} 
\includegraphics[width=0.33\columnwidth]{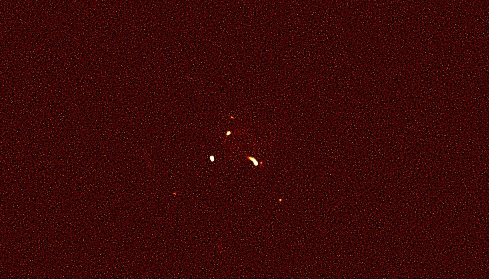}
\includegraphics[width=0.33\columnwidth]{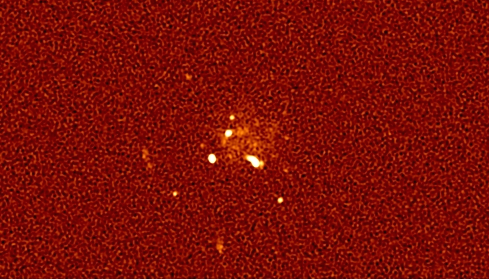}
\includegraphics[width=0.33\columnwidth]{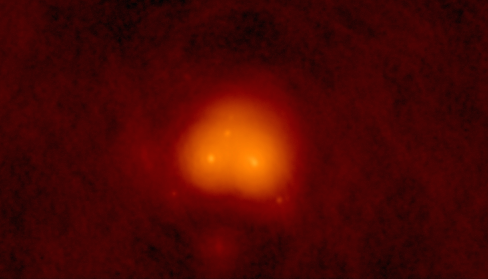}
\includegraphics[width=0.33\columnwidth]{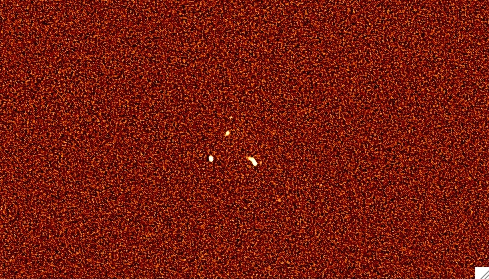}
\includegraphics[width=0.33\columnwidth]{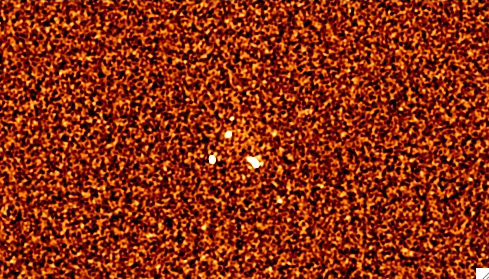}
\includegraphics[width=0.33\columnwidth]{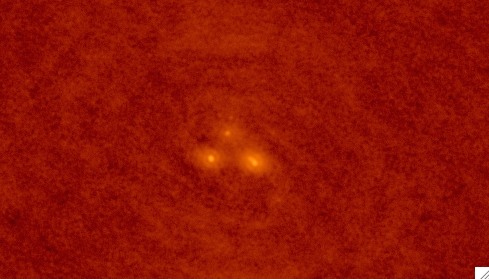}
\caption{Dirty maps resulting from simulated observations of the model cluster at z=0.5 for {\em Cases A}, {\em B} and {\em C} (from {\em left} to {\em right}). {\em Top} panels show results for SKA1-MID, {\em bottom} panel for SKA1-SUR.}
\label{fig:dirty}
\end{figure*}

We have performed different imaging tests both for SKA1-SUR and SKA1-MID with w-projection correction included and no other wide-field effects simulated. In the following we will present results for: {\em Case A)} a uniform weighting scheme with 1 arcsec taper; {\em Case B)} a uniform weighting scheme with 5 arcsec taper; {\em Case C)} natural weighting. Examples of the resulting dirty maps for the cluster at z=0.5 are shown in Fig.\,\ref{fig:dirty}. Despite the fact that nearly no convolution artefacts are present in the images, at the highest resolution ({\em Case A}) only the brightest radio sources are visible and the diffuse radio emission is completely below the noise. The diffuse emission of the radio halo is instead already distinguishable on the dirty map in {\em Case B}. 

Note that our approach, based on simulated observations, differs from the feasibility study presented by \citeauthor{cassano15}, this Volume. In that case, similarly to \citet{ferrari11}, the authors use a criterion based on a threshold in surface brightness to estimate if a radio halo of a given luminosity can be detected by SKA. For this, they need to assume a certain, generalised surface brightness profile for radio halos (e.g. \cite{cassano15} assume that about half of the total halo flux is contained in about half halo radius, while \citet{ferrari11} adopts a brightness profile as a function of radius derived from \citet{govoni01}).

\section{New deconvolution and source detection method} \label{moresane}

\begin{table*}[t]
  \centering{     
\begin{tabular}{c c c c c c c}
\hline
\noalign{\vskip 0.1cm}    
 {\em Cases} & {\em A} at z=0.5 & {\em B} at z=0.5 & {\em C} at z=0.5 & {\em A} at z=1.0 & {\em B} at z=1.0 & {\em C} at z=1.0 \\
\noalign{\vskip 0.1cm}    
\hline 
\hline                       
\noalign{\vskip 0.1cm}    
{\bf SKA1-MID} & & & & & & \\
\noalign{\vskip 0.1cm}    
\hline
\noalign{\vskip 0.1cm}    
 Resolution [arcsec] & 1.8 & 4.5 & 10.4 & 1.8 & 4.5 & 10.4\\      
\noalign{\vskip 0.1cm}    
\hline
Sensitivity [$\mu$Jy/beam] & 2.4 & 1.7 & 0.8 & 2.5 & 1.9 & 1.2\\
\noalign{\vskip 0.1cm}    
\hline   
\hline                        
\noalign{\vskip 0.1cm}    
{\bf SKA1-SUR} & & & & & & \\
\noalign{\vskip 0.1cm}    
\hline
\noalign{\vskip 0.1cm}    
 Resolution [arcsec] & 1.8 & 4.9 & 7.4 & 1.8 & 4.9 & 7.4\\      
\noalign{\vskip 0.1cm}    
\hline
Sensitivity [$\mu$Jy/beam] & 6.9 & 7.0 & 5.6 & 6.5 & 7.0 & 6.0\\
\noalign{\vskip 0.1cm}    
\hline   
\hline                        
\noalign{\vskip 0.1cm}   
\end{tabular}
}
\caption{Final resolution and rms sensitivity of the restored maps obtained with a total observing time of 8 hours and using {\tt MS-CLEAN} (see Sect.\,2.2 for more details).}
\label{table1} 
\end{table*}

\begin{figure*} 
\centering
\includegraphics[width=0.85\columnwidth]{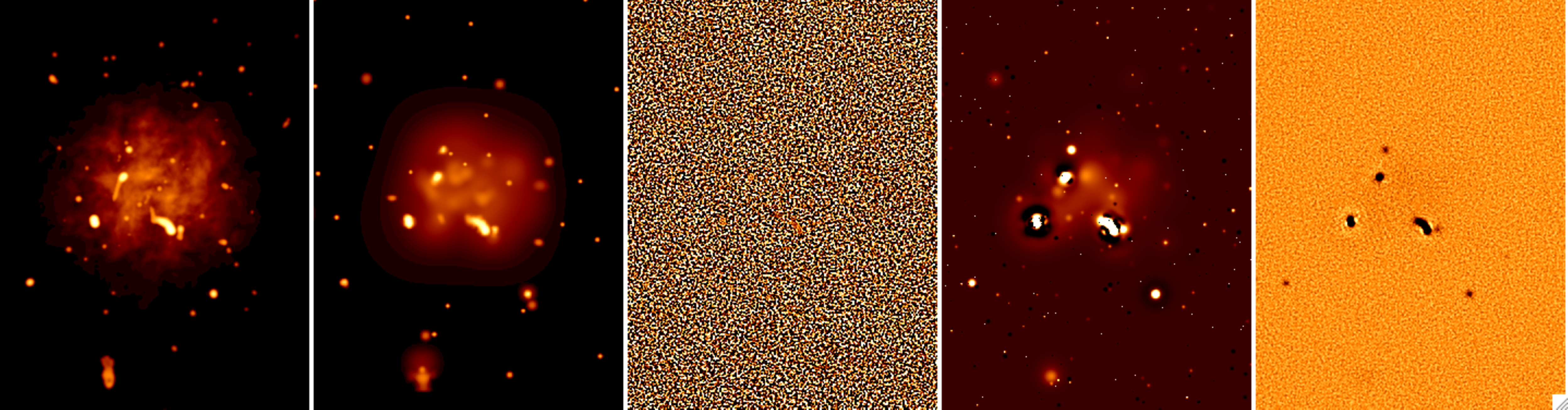}
\includegraphics[width=0.85\columnwidth]{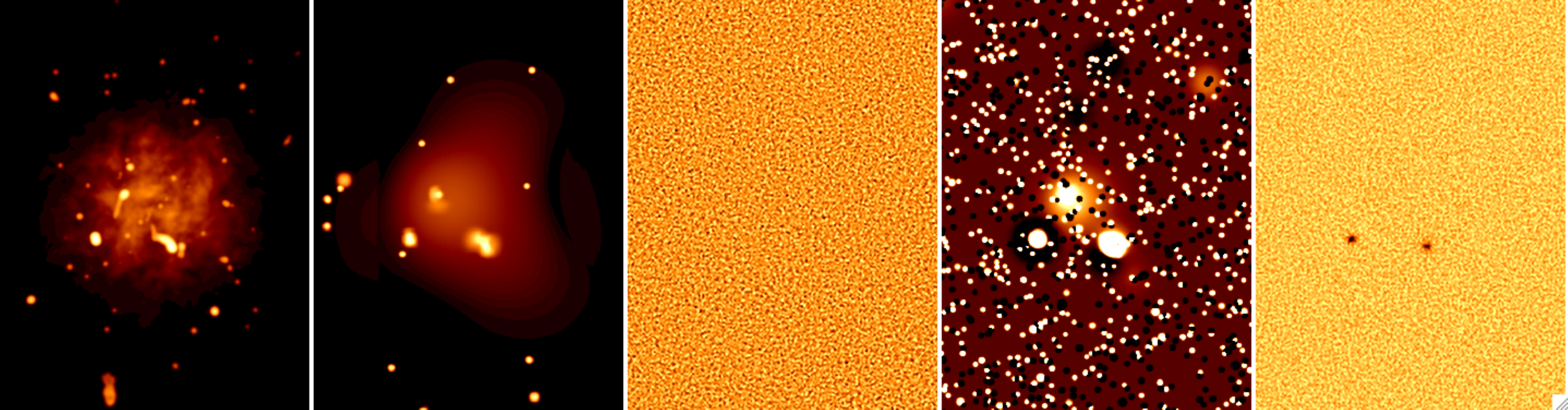}
\caption{Results of deconvolution for the model cluster at z=0.5 ({\em top}) and z=1.0 ({\em bottom}) observed with SKA1-MID and adopting the imaging parameters of {\em Case A}. From {\em left} to {\em right}: model cluster map convolved at the same resolution of simulated observations; source model resulting from {\tt MORESANE} deconvolution algorithm convolved at the same resolution of simulated observations; {\tt MORESANE} maps of residuals; {\tt MS-CLEAN} components convolved at the same resolution of simulated observations; {\tt MS-CLEAN} maps of residuals. The model images are saturated at the same level.}
\label{fig:results1}
\end{figure*}

\begin{figure*} 
\centering
\includegraphics[width=0.85\columnwidth]{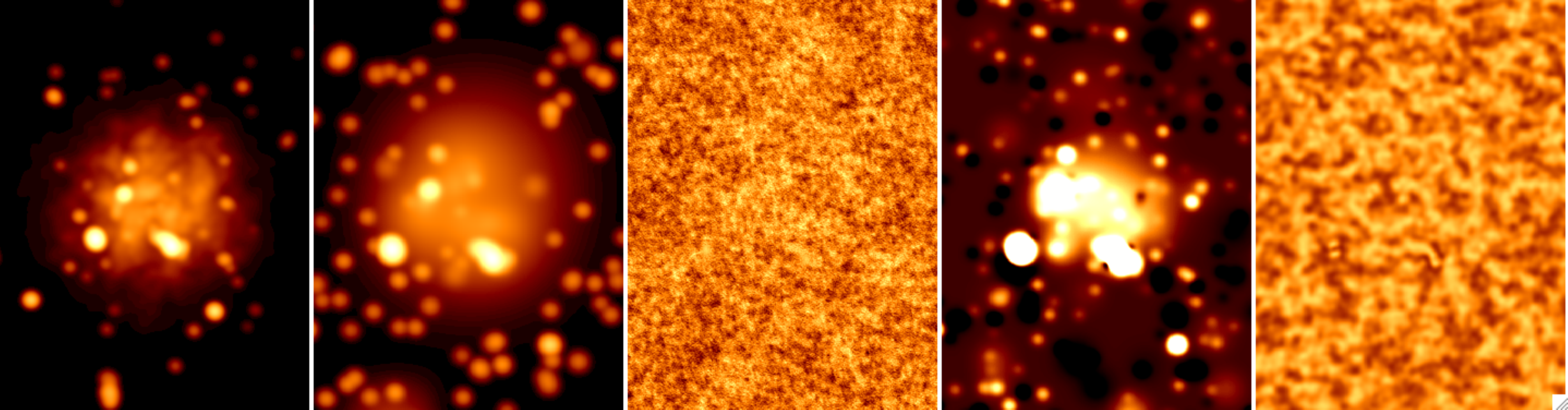}
\includegraphics[width=0.85\columnwidth]{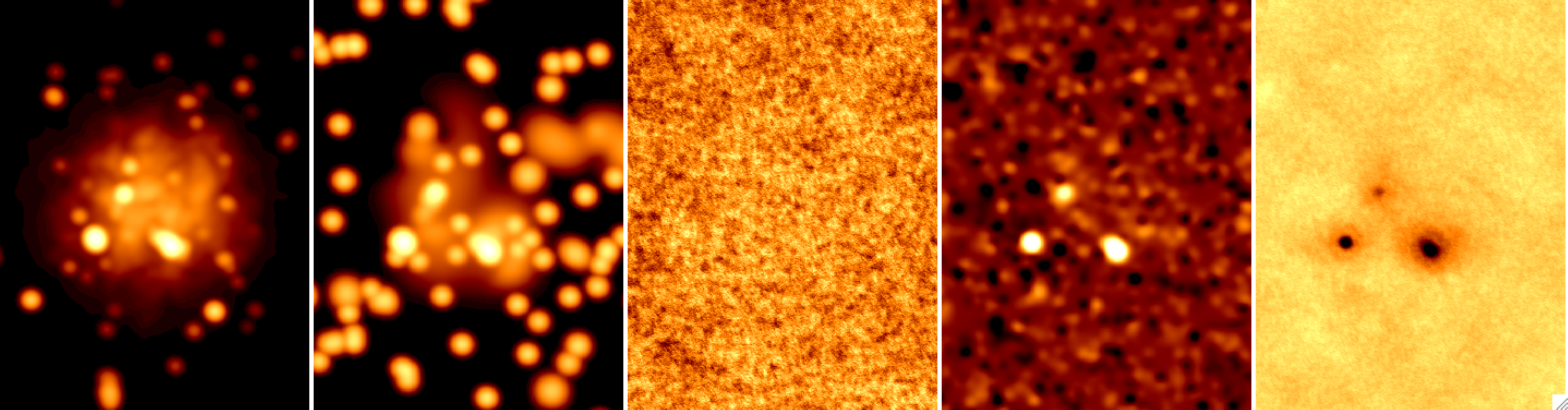}
\caption{Results of deconvolution for the model cluster at z=0.5 ({\em top}) and z=0.7 ({\em bottom}) observed with SKA1-SUR and adopting the imaging parameters of {\em Case C}. Columns are the same as in Fig.\,3.}
\label{fig:results2}
\end{figure*}

We run both H\"ogbom and Multi-Scale ({\tt MS-}{\tt CLEAN})\footnote{Eigth scales are used for {\tt MS-}{\tt CLEAN}: [0,2,4,8,16,32,64,128].} algorithms \citep{hogbom74,cornwell08} on the dirty maps down to 2$\sigma$ level. In both cases, we use the {\tt lwimager} software implemented in {\tt MeqTrees}, a stand-alone imager based on the {\tt CASA} libraries and providing {\tt CASA}-equivalent implementations of various {\tt CLEAN} algorithms. No {\tt CLEAN} boxes are used in our tests, since we aim at verifying results for a fully automatic data reduction (as required for SKA). The rms sensitivity and final resolution (determined by a combination of the adopted taper, weighting and uv-coverage) of the restored maps both for SKA1-SUR and SKA1-MID tests are given in Table\,\ref{table1}. The {\tt MS-CLEAN} components (convolved at the same resolution of simulated observations) and maps of residuals are shown in the fourth and fifth columns of Figs.\,\ref{fig:results1} and \ref{fig:results2}.

We then run the deconvolution algorithm {\tt MORESANE} \citep{dabbech12, dabbech14} on dirty maps. {\tt MORESANE}, whose results are shown in the second and third columns of Figs.\,\ref{fig:results1} and \ref{fig:results2}, belongs to the family of new algorithms based on the theory of Compressed Sensing \citep[see Sect. 4.4.2 in][]{norris13}. More specifically, {\tt MORESANE} allies complementary types of sparse image models \citep{dabbech14}. The clear advantage is that its reconstructed image allows the detection of both extended and compact radio sources, reproducing in a accurate way their morphologies (see Fig.\,\ref{fig:results1}). Further tests indicate that flux measurements can be derived as a direct output of {\tt MORESANE}, since the photometry of the input model sky is conserved in the deconvolved image \citep{dabbech14}. The algorithm has been conceived and optimised for the detection and characterisation of very low-surface brightness and extended radio sources, resulting in the case presented here in the non-trivial detection of the very weak radio halo as well as in the good recovery of tailed radio galaxy morphologies. In addition, contrarily to {\tt CLEAN}, the contamination by fake model components has been proven to be extremely weak, when not absolutely zero. Based on all these elements, the output of {\tt MORESANE} can therefore be used for source catalog purposes \citep[see][for a detailed description of {\tt MORESANE} and a quantitative comparison with other existing deconvolution methods]{dabbech14}.

Thanks to the image reconstructed by {\tt MORESANE}, we can conclude that 8 hour observations with SKA1-MID will allow us to easily detect the different components of our model cluster (from tailed radio galaxies to the low surface brightness radio halo) up to z=1 and with an excellent resolution ($\sim$ 1 arcsec, {\em Case A}). Through 8 hours of observation with SKA1-SUR, we are instead able to get hints of the possible presence of a diffuse radio source up to z $\sim$ 0.7 only when adopting a higher sensitivity natural weighting scheme ({\em Case C}).

\subsection{Notes on SKA1-LOW}

Diffuse intracluster radio sources are generally characterised by steep synchrotron spectra. This, together with their low-surface brightness and the possible spectral steepening at high radio frequencies due to electron ageing, make them more easily detectable at long wavelengths. In addition, a unique prediction of turbulence acceleration models is the existence of ultra-steep radio halos, not associated to major cluster mergers, but to less energetic merging events \citep{cassano13}. Low-frequency observations are required to detect this kind of sources, as well as old population of electrons, for instance in dying or re-started radio galaxies at the center of galaxy clusters.

With a maximum baseline of 100 km, we can expect a maximum resolution of about 5 arcsec at 150 MHz and about 9 arcsec at 70 MHz, resulting in surface brightness confusion levels of the order of 140 and 245 nJy/arcsec$^2$ \citep[see Fig.\,4 in][courtesy J. Condon]{ferrari13}\footnote{Assuming a confusion noise that scales proportionally to $\nu^{-0.7}$, as a typical radio source.}. This imposes a much more severe limit in the sensitivity of SKA1-LOW  to diffuse emission from clusters compared to, for instance, the lowest frequency part of SKA1-MID \citep[see Fig. 6 in][]{ferrari13}. Higher resolution SKA1-LOW observations could not only allow us to achieve a higher sensitivity to diffuse radio emission by removing point sources and by re-imaging at lower resolution the subtracted data \citep[see e.g.][]{vazza15}, but are also absolutely required to discriminate between the radio emission from active galaxies and from diffuse intra-cluster radio sources, particularly at high-redshift (z$\gtrsim$1). Current resolutions achieved by SKA1-LOW are therefore limiting low-frequency high/intermediate-z cluster science in Phase 1.

\section{Conclusions and future plans} \label{conclusion}

In this work, based on simulated SKA1 observations of galaxy clusters, we show that prospects are good for the study of non-thermal cluster physics, in particular thanks to new developments in the deconvolution and source detection steps that are here optimised for the analysis of extended and diffuse radio sources. Note that, in our simulations, we adopt a narrow band-width (50 MHz, Sect.\,\ref{simulations}). The quality of the results indicate that we will be able to get multi-frequency images of diffuse cluster radio sources within each of the large SKA1 bands, thus enabling detailed spectral index studies of galaxy clusters, an essential tool for our understanding of their NT physics \citep[e.g.][]{orru07}.

Based on the results highlighted in Sect.\,\ref{simulations}, we conclude that SKA1-MID is an extremely powerful   instrument for radio analyses of galaxy clusters: relatively deep ($\lesssim$8 hours) follow-up observations of interesting targets from multi-wavelength (optical, X-ray, Sunyaev-Z'eldovich, ..., see the Chapter by \citeauthor{grainge15}, this Volume) cluster catalogs can allow detailed studies of both tailed radio galaxies and relatively low-luminosity ($P_{\rm 1.4~GHz} \approx 10^{24}$ W/Hz, see Sect.\,\ref{model}) diffuse radio halos up to at least z=1.0. A 2-years all-sky survey with SKA1-SUR can provide a completely independent interesting catalog of new candidates of diffuse cluster sources up to z$\sim$0.7, to be possibly followed up with SKA1-MID.
 
\begin{SCfigure} 
\centering
\caption{Frequency {\it vs.} largest angular scale (LAS) detectable as a function of three different array minimum baselines (${\rm B}_{\rm min}$). The angular scale corresponding to a typical 1 Mpc intra-cluster radio source at different redshifts is also indicated (we assume a $\Lambda$CDM cosmology with $\Omega_{\Lambda} = 0.7$ and $\Omega_{\rm M} = 0.3$).}
\includegraphics[width=0.65\columnwidth]{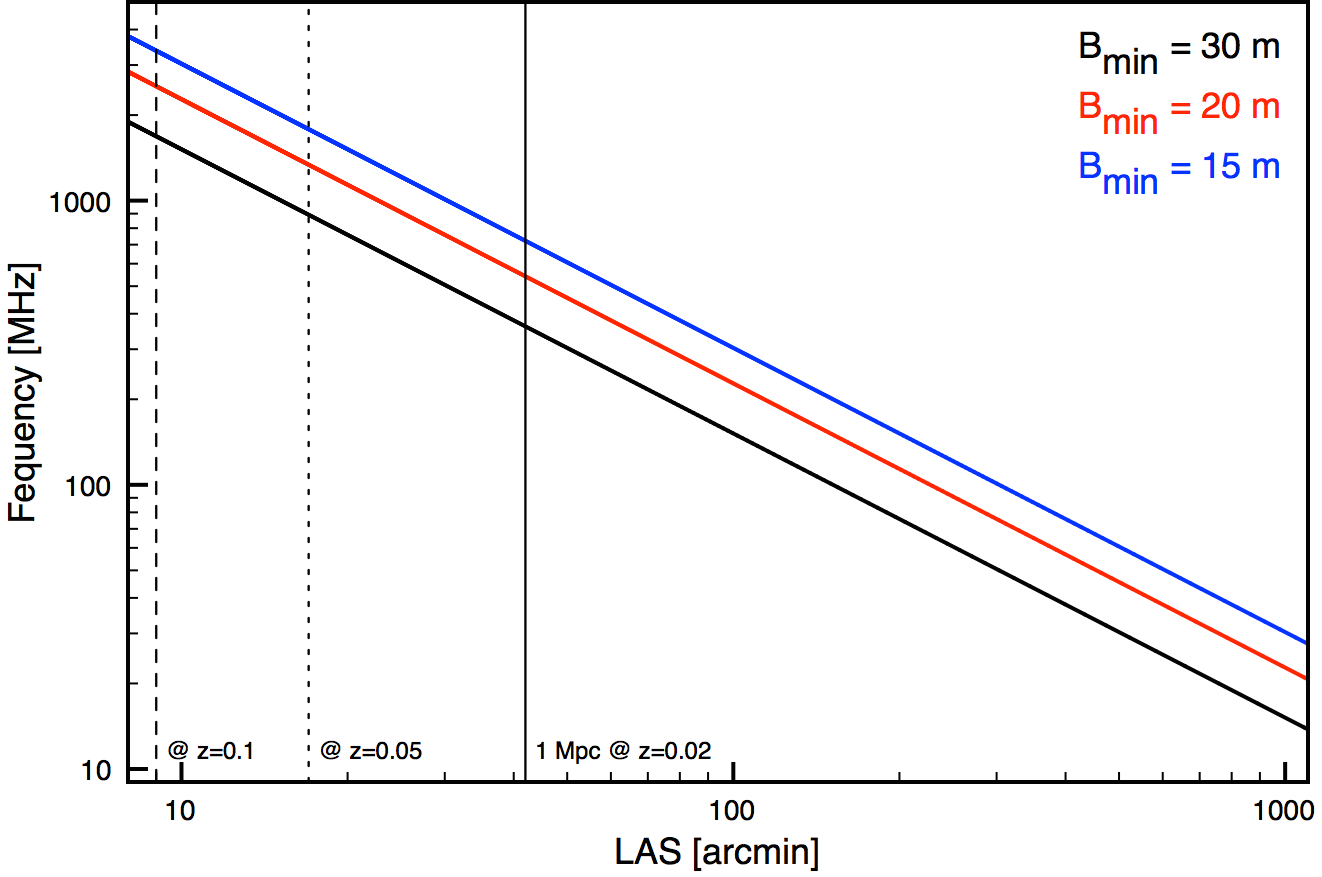}
\label{fig:las}
\end{SCfigure}

On a final cautionary note, Fig.\,\ref{fig:las} shows the largest angular scale (LAS, in arcmin) that can be detected as a function of the observed frequency and minimum array baseline. For reference, we show the angular scale corresponding to a typical 1 Mpc intra-cluster radio source at different redshifts. We can note, for instance, that, at 1.4 GHz, a minimum baseline of approximately 20-30 m does not allow to detect structures larger than $\sim$ 1 Mpc at z $<$ 0.05. In order to image giant radio sources down to very low--redshifts, we can either combine single-dish and interferometric data to completely fill in the gap down to 0m-spacing or perform coherent mosaicking observations by scanning the interferometer over the extended source with a regular (at least Nyquist--spaced) grid \citep[see e.g.][and references therein]{holdaway99}.

\vspace{0.2cm}

The analysis developed in this paper will be extended in future works, in particular:

\begin{itemize}

\item the detectability with SKA1 of simulated radio relics and radio bubbles \citep[e.g.][]{vazza12b,roediger07} will be investigated and compared to similar feasibility studies for SKA precursors and pathfinders (JVLA, LOFAR, GMRT, MeerKAT, ASKAP, \dots). Polarisation studies for targeted observations might also be included;

\item due to the importance of low-frequency observations for cluster science, a more extended feasibility study will be taken into account for SKA1-LOW; 

\item on longer time-scale, we would like to develop similar observational simulations for the Phase 2 configuration of the SKA array. At present, we are limited by computer resources for performing this part of the work.

\end{itemize}

\vspace{0.2cm}

\noindent {\it Acknowledgements}

\noindent We thank the referees of this paper for their useful comments and the SKA office for the conference organisation. We acknowledge financial support by the ``{\it Agence Nationale de la Recherche}'' through grant ANR-09-JCJC-0001-01, the ``{\it Programme National Cosmologie et Galaxies (2014)}'', the {\it BQR} program of Lagrange Laboratory (2014), the ``PHC PROTEA'' programme (2013), the joint doctoral program ``{\it r\'egion PACA-OCA}'' (2011). S. Makhathini acknowledges financial support from the National Research Foundation of South Africa. O. Smirnov's research is supported by the South African Research Chairs Initiative of the Department of Science and Technology and National Research Foundation.

\end{document}